\documentclass[aps,prl,twocolumn,showpacs,superscriptaddress]{revtex4-1} 

\usepackage{graphicx}  
\usepackage{dcolumn}   
\usepackage{bm}        
\usepackage{amsmath}   
\usepackage{amsfonts}
\usepackage{amssymb}
\usepackage{natbib}
\usepackage[percent]{overpic}
\usepackage{color}
\definecolor{darkblue}{rgb}{0,0,0.6}
\definecolor{darkred}{rgb}{0.6,0,0}
\definecolor{darkgreen}{rgb}{0,0.6,0}
\definecolor{purple}{rgb}{0.6,0,0.6}
\usepackage{hyperref}
\hypersetup{
   colorlinks=true,       
   linkcolor=black,         
   citecolor=blue,        
   urlcolor=blue           
}

\newif\iffigures
\figurestrue 

\newcommand \gdot {\dot{\gamma}}
\newcommand \df {d_{\mathit{f}}}

\begin{document}

\title{Driving rate dependence of avalanche statistics and shapes at the yielding transition}

\author{Chen Liu}
\affiliation{Universit\'e Grenoble Alpes, LIPHY, F-38000 Grenoble, France}
\affiliation{CNRS, LIPHY, F-38000 Grenoble, France}

\author{Ezequiel E. Ferrero}
\affiliation{Universit\'e Grenoble Alpes, LIPHY, F-38000 Grenoble, France}
\affiliation{CNRS, LIPHY, F-38000 Grenoble, France}

\author{Francesco Puosi}
\affiliation{Ecole Normale Sup\'erieure de Lyon, Laboratoire de Physique CNRS, 46 all\'ee d'Italie, 69364 Lyon Cedex 7, France}
\affiliation{CNRS, LIPHY, F-38000 Grenoble, France}

\author{Jean-Louis Barrat}
\affiliation{Universit\'e Grenoble Alpes, LIPHY, F-38000 Grenoble, France}
\affiliation{CNRS, LIPHY, F-38000 Grenoble, France}

\author{Kirsten Martens}
\affiliation{Universit\'e Grenoble Alpes, LIPHY, F-38000 Grenoble, France}
\affiliation{CNRS, LIPHY, F-38000 Grenoble, France}


\begin{abstract}
We study stress time series caused by plastic avalanches in athermally sheared disordered materials.
Using particle-based simulations and a mesoscopic elasto-plastic model, 
we analyze size and shear-rate dependence of the stress-drop durations and size distributions together with their average temporal shape.
We find critical exponents different from mean-field predictions, and a clear asymmetry for individual avalanches.
We probe scaling relations for the rate dependency of the dynamics and we report a crossover towards mean-field results for strong driving.
\end{abstract}

\pacs{62.20.F-, 45.70.Ht, 63.50.Lm, 64.60.av }

\maketitle

Many materials respond to slow driving with strongly intermittent dynamics.
Examples include Barkhausen noise in ferromagnets~\cite{BarkhausenPhysikZ1917,DurinPRL2000,DurinBookChapter2006},
stick-slip motion in earthquakes~\cite{Ruina1983}, serration dynamics in plasticity of solids~\cite{DasturMetallTransA1981},
and avalanche dynamics in crack propagation~\cite{BonamyPRL2008,LaursonPRE2010}, driven foams~\cite{CantatPhysFluids2006} 
and domain wall motion~\cite{RepainEPL2004}.

As in equilibrium critical phenomena, global quantities linked to 
such bursting collective events are usually power law distributed
and allow for the introduction of scaling functions.
In the slow driving limit, the onset of motion can
be interpreted as an out-of-equilibrium phase transition,
suggesting the existence of families 
of systems that display similar avalanche statistics.
To better identify this universality classes,
both experimental~\cite{ChrzanPRB1994,SpasojevicPRE1996,KuntzPRB2000,ZapperiNatPhys2005,LaursonPRE2006,
PapanikolauNatPhys2011,LaursonNatCom2013,AntonagliaPRL2014} and
theoretical~\cite{SethnaNat2001,MehtaPRE2002,LeDoussalPRE2014,ZapperiNatPhys2005, ZhaoJPhys2014} works have
discussed the avalanche ``shapes'', going beyond the study of scaling exponents.

In deformation experiments of amorphous systems, such as grains, 
foams or metallic glasses, avalanche dynamics
are typically evidenced in the time series of the deviatoric component of the stress tensor.
In the limit of vanishing deformation rate 
we approach the so-called ``yielding transition''.
The question whether yielding can be characterized as a continuous dynamical phase transition,
belonging to a specific universality class, is still under debate.
The analysis of avalanche statistics close to yielding has therefore a
particular relevance.

In this letter, we study the emerging yielding dynamics in 
a simple shear geometry with imposed driving rate.
Our focus lies on the shear-rate dependence of the avalanche statistics and thus 
complement recent quasi-static studies~\cite{SalernoMaRoPRL2012,SalernoRoPRE2013,BudrikisPRE2013,LinPNAS2014},
To address the low shear-rate regime we use a coarse-graining approach, proven to yield qualitative 
and quantitative relevant predictions~\cite{Picard2005,Rodney2011,MartensPRL2011,Talamali2012,NicolasPRL2013,FerreroPRL2014},
and compare the low shear-rate results of our meso-scale model 
with quasistatic particle-based simulations.

\noindent
{\it Molecular dynamics (MD) --}
We consider a mixture of $\tt A$ and $\tt B$ particles interacting
via a Lennard-Jones potential:
$V_{\tt AB}(r)=4\epsilon_{\tt AB}[({\sigma_{\tt AB}}/{r} )^{12}-({\sigma_{\tt AB}}/{r})^{6}]$
with $r$ being the distance between two particles.
Units of energy, length and mass are defined by $\epsilon_{\tt AA}$, $\sigma_{\tt AA}$ and $m_{\tt A}$;
the unit of time is given by $\tau_0=\sigma_{\tt AA}\sqrt{(m_{\tt A}/\epsilon_{\tt AA})}$.
The potential is truncated at $R_c=2.5$ and a force smoothing is
applied between an inner cut-off $R_{in}=2.2$ and $R_c$.
The two species of particles have equal mass $m$, but different interaction parameters to prevent crystallization.
We set $\epsilon_{\tt AA}=1.0 $, $\epsilon_{\tt AB}=1.5 $, $\epsilon_{\tt BB}=0.5 $, $\sigma_{\tt AA}=1.0$,
$\sigma_{\tt AB}=0.8$ and $\sigma_{\tt BB}=0.88$ and $m=1$.
The ratio of particles of species $\tt A$ and $\tt B$ is chosen $N_{\tt A}/N_{\tt B}=13/7$ and $8/2$
for $2d$ and $3d$ systems, respectively.
Glassy states are obtained (with LAMMPS~\cite{PlimptonJCP1995}) by quenching to zero temperature at constant volume systems equilibrated at $T=1$.
An athermal system is achieved by applying to each particle a viscous drag force
$\mathbf{F}_{drag}=-\Gamma\mathbf{v}$, where $\mathbf{v}$ is the particle peculiar velocity.
We condition the dynamics to be strongly overdamped~\cite{SalernoMaRoPRL2012,PuosiPRE2014} ($\Gamma=1$).
Avalanche statistics are obtained following a quasistatic protocol~\cite{SalernoMaRoPRL2012,SalernoRoPRE2013}.
We impose simple shear at rate $\gdot=10^{-6}$ by deforming the box dimensions and remapping the particle positions.
Following~\cite{SalernoMaRoPRL2012}, the shear-rate $\gdot$ is set to zero when a steep increase in kinetic energy occurs 
(onset of plastic deformation) and only restored when the kinetic energy drops below a
threshold.

\noindent
{\it Elasto-plastic (EP) model--}
We coarse-grain an amorphous medium onto a mesoscopic lattice: each node represents a block of material
holding exactly one shear transformation~\cite{ArgonAM1979,TanguyEPJE2006,MaloneyPRE2006,PuosiPRE2014},
for which we assume the same geometry as the globally applied simple shear.
To each site $i$ we associate a local scalar shear stress $\sigma_i$ and a state
variable $n_i$, indicating whether the site plastically deforms ($n=1$) or not ($n=0$).
Local stresses evolve with the overdamped dynamics:
\begin{equation}\label{eq:Eqofmotion}
  \partial_t\sigma_i = \mu\gdot + \mu \sum_{j} G_{ij} \partial_t \gamma^{pl}_j
\end{equation}
with $\mu=1$ the elastic modulus, $\gdot$ the externally applied shear-rate, $\tau=1$ a mechanical relaxation time and
$\partial_t\gamma^{pl}_j=\frac{n_{j}\sigma_{j}}{\mu\tau}$ the strain rate produced by a plastic rearrangement at site $j$.
$G_{ij}$ denotes the discretized Eshelby propagator~\cite{Eshelby}, that obeys a quadrupolar symmetry in the shear plane with
a dipolar long-range character, $G(\bm{r},\bm{r'})=\cos(4\theta_{\bm{r}\bm{r'}})/|\bm{r}-\bm{r'}|^d$.
A site yields ($n_i=0 \rightarrow 1$) when its stress reaches a local threshold
$\sigma_i \geq \sigma^y_i$, and recovers its elastic state ($n_i=1 \rightarrow 0$) when a prescribed
local deformation increment is attained after yielding,
$\int | \partial_t \sigma_i/\mu + \partial_t\gamma^{pl}_i|dt \geq \gamma_c$.
Each time a site yields a new yield stress $\sigma^y_i$ is drawn from a distribution of mean $\sigma_0$.
Model details and parameter choices can be found in Ref.\cite{NicolasEPL2014} and in the Supplemental Material~\cite{SM}.

\noindent
{\it Stress-drop statistics and shear-rate dependence--}
From the stress-time series we individualize stress-drops,
and define an extensive quantity $S$ proportional to the absolute
stress difference multiplied by the system volume.
We compare in Fig.\ref{fig:PofSfor2Dand3DandMD} the stress-drop distributions $P_S$ in the limit of low $\gdot$
for the EP model with the quasistatic MD results.
In both two ($2d$) and three dimensions ($3d$), apart from a plateau regime for
small stress-drops that depends on shear-rate, numerical integration step and system size, we 
fit the data using a power-law $P_S\sim S^{-\tau}f(S/S_c)$, with
$f$ an exponentially decaying cut-off function \cite{PlanetPRL2010} (exponent definitions in Table~\ref{table:criticalexponents}).
Noticing that the distributions $P_S$ become independent of $\dot{\gamma}$ in the zero shear-rate limit
and in agreement with previous works~\cite{SalernoRoPRE2013,LinPNAS2014},
we postulate a system size dependent cut-off $S_c \sim L^{\df}$,
with $\df$ the \textit{fractal dimension} of the avalanches~\cite{BaileyPRL2007, SalernoRoPRE2013, LinPNAS2014}.
The comparison of these stress-drop statistics with MD results reveals 
a fair agreement, up to an arbitrary scaling factor related to the difference in simulated length scales.

\iffigures
\begin{figure}[t!]
\begin{center}
\includegraphics[width=\columnwidth, clip]{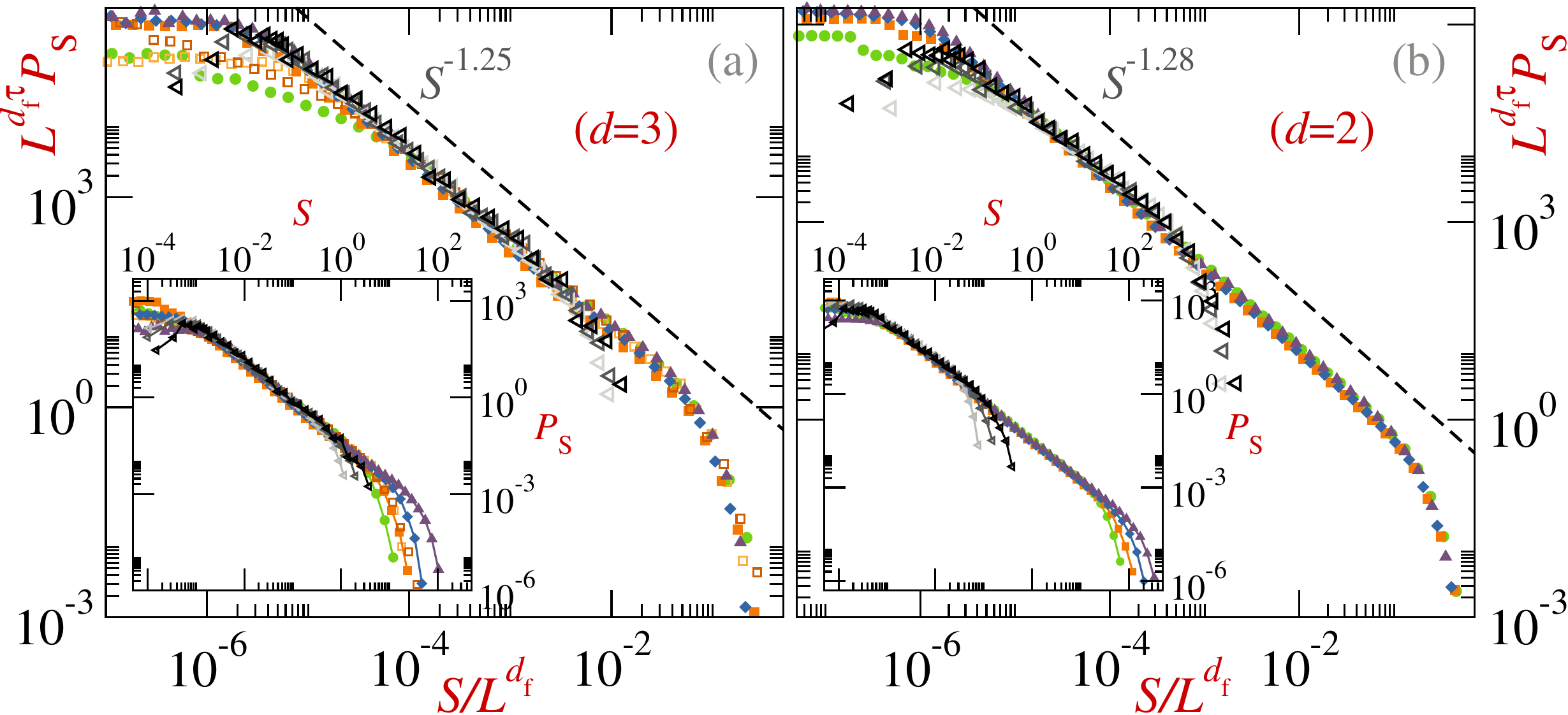}
\end{center}
\caption{\textit{Stress-drop size distributions}.
Main panels show rescaled distributions $L^{\df\tau}P_S$ vs. $S/L^{\df}$ of the EP model compared to
MD quasistatic simulations (arbitrary shift applied for the comparison).
Insets show not-scaled curves.
(a) 3$d$EP model data for linear system-sizes $L=16$ (green circles), $32$ (orange squares), $64$ (blue diamonds), $128$ (plum triangles)
and shear-rate $10^{-4}$ (full symbols).
For $L=32$, $\gdot=10^{-3},10^{-5}$ are also shown (light and dark orange open squares). 
Gray scale triangles correspond to quasistatic $3d$MD with $L=40,60,80$ (from light to dark).
(b) 2$d$EP data for linear system-sizes $L=256$ (green circles), $512$ (orange squares), $1024$ (blue diamonds) and $2048$ (plum triangles)
at $\gdot=10^{-5}$.
Gray scale triangles correspond to quasistatic $2d$MD with $L=80,160,320$ (from light to dark).
} 
\label{fig:PofSfor2Dand3DandMD}
\end{figure}
\fi

The fitted values of $\tau$ for the EP model, both in two and three dimensions
($\tau_{\tt 2d}\simeq 1.28$, $\tau_{\tt 3d}\simeq 1.25$),
compare very well with our and earlier obtained MD results~\cite{SalernoMaRoPRL2012, SalernoRoPRE2013}, are compatible with previous
lattice models~\cite{TalamaliPRE2011}, and lie within error bars of those provided by
FEM models~\cite{SandfeldJSTAT2015}.
Still, they disagree with what was obtained with quasistatic protocols in cellular automaton
models~\cite{LinPNAS2014} (especially in 3$d$ where $\tau^{\tt QS}_{\tt 3d} \simeq 1.43$),
and they contrast even more with the usual mean-field prediction~\cite{DahmenNatPhys2011} $\tau^{\tt MF}=3/2$
(see~\cite{Jagla_arXiv2015} for an alternative analysis).
The values obtained for $\df$ ($\df^{\tt 2d}\simeq 0.9$, $\df^{\tt 3d}\simeq 1.3$)
are compatible with quasistatic MD simulations, 
but slightly smaller than those reported in automaton models~\cite{LinPNAS2014}.
They suggest a line geometry of the correlated slip events~\cite{LemaitrePRL2009, BudrikisPRE2013},
with a modest but clear trend towards a more compact structure in 3$d$. 

\iffigures
\begin{figure}[t!]
\begin{center}
\includegraphics[width=\columnwidth, clip]{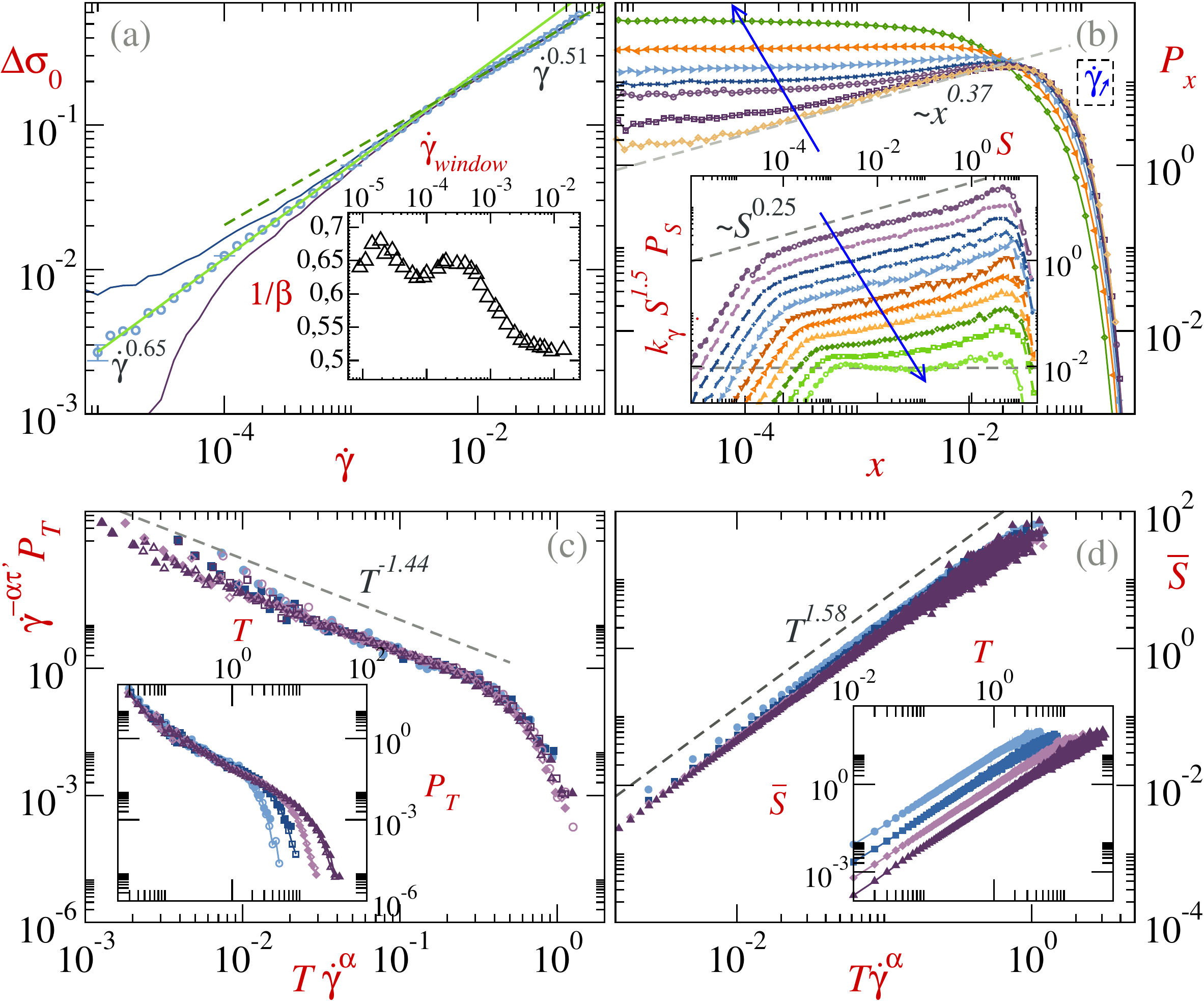}
\end{center}
\caption{\textit{Shear-rate dependency of the dynamics for the 3$d$EP model}.
(a) Log-log plot of $\Delta\sigma_0\equiv(\sigma-\sigma_c)/\sigma_0$ vs $\gdot$.
Circles correspond to the best estimation of $\sigma_c/\sigma_0 = 0.687$;
alongside lines, to choices of $0.683$ and $0.691$ instead.
Full and dashed lines are power-law fits in selected ranges (extrapolated for comparison).
Inset: Crossover of $1/\beta$ as explained in the text.
(b) Steady-state distributions $P_x$ of the local distances to threshold $x\equiv\sigma_y-\sigma$ for different
shear-rates $\gdot\in\{10^{-1.4},\ldots,10^{-5}\}$.
Inset: Stress-drop distributions for $\gdot\in\{ 10^{-1},\ldots,10^{-3}\}$, rescaled and shifted
as explained in the text.
Arrows indicate the sense of increasing shear-rate.
(c) Rescaled distributions of stress-drop duration $\gdot^{-\alpha\tau'}P_T$ vs. $T\gdot^{\alpha}$ for
$\gdot=10^{-2},10^{-3},10^{-4},10^{-5}$ (from light blue to dark plum, left to right in inset),
and system-sizes $L=64$ (closed symbols) and $128$ (open symbols).
The dashed line shows a law $P_T\sim T^{-1.44}$.
Inset: Unscaled data.
(d) Average size $\bar{S}$ for stress-drops of the same duration as a function of $T\gdot^\alpha$
for $L=64$ and $\gdot=10^{-2},10^{-3},10^{-4},10^{-5}$.
The dashed line shows $\bar{S}\sim T^{1.58}$.
Inset: Unscaled data, shear-rate decreases from left to right.
} 
\label{fig:FlowcurvePofXSofTPofT}
\end{figure}
\fi

Some main results concerning the finite driving rate are summarized in Fig.~\ref{fig:FlowcurvePofXSofTPofT}
for the $3d$EP model, similar results are found for the $2d$ case (not shown).
The consequences of applying a finite shear-rate are twofold\footnote{To extrapolate our results we
assume that the yielding transition is continuous.}:

(I) The first important observation is that with increasing driving rate 
the critical exponents tend towards the mean-field predictions.
The yielding exponent $\beta$ for example, defined through $\gdot \propto (\sigma-\sigma_c)^\beta$,
can be derived from the fits in Fig.\ref{fig:FlowcurvePofXSofTPofT}(a)
rendering a non-trivial value $\beta \simeq 1.55$ in the low shear-rate regime.
For larger shear-rates this value crosses over to $\beta \sim 2$ predicted by the Hebraud-Lequeux
model~\cite{HebraudPRL1998}. 
By sliding a fixed size logarithmic window in $\gdot$ (comprising $\sim$12 points of
the main plot data set) and fitting within, we show the resulting $1/\beta$ as a
function of the starting position of the window in the inset of Fig.~\ref{fig:FlowcurvePofXSofTPofT}(a).
Similarly we observe a crossover of the exponents in the steady-state distribution $P_x$ of the local stress
excess~\cite{LinEPL2014,MullerARCMP2015} $x_i\equiv\sigma^y_i-\sigma_i$, Fig.\ref{fig:FlowcurvePofXSofTPofT}(b).
Again in the limit of vanishing shear-rates we observe the curves approaching a shape that initially
grows as $P_x \sim x^\theta $ with a non-trivial exponent, as found in the quasistatic
case~\cite{LinEPL2014, LinPNAS2014}, attributed to an anomalous random walk process of the local stress
with an absorbing boundary condition at $x=0$~\cite{LinarXiv2015}.
However, as we increase the shear-rate
$P_x$ changes, eventually yielding $\theta \simeq 0 $.
The driving progressively dominates over the signed kicks
from elastic interactions, yielding a biased diffusion of the $x$'s values. 
This ultimately produces a strictly positive local stress evolution,
resembling the $x$ dynamics of the depinning problem~\cite{LinPNAS2014}.
The inset of Fig.\ref{fig:FlowcurvePofXSofTPofT}(b) shows a feature compatible with the shear-rate dependence
of $P_x$ and with the $\beta$ crossover.
For different shear-rates, we plot $k_{\gdot} S^{1.5} P_S$ vs. $S$, where $k_{\gdot}$ is an arbitrary
scaling coefficient to separate the curves and improve visualization.
We observe a range of low shear-rates where the slope of the transformed distributions is almost unchanged
and fully consistent with Fig.\ref{fig:PofSfor2Dand3DandMD}(a).
Above a rate of deformation of about $\sim0.015$, curves progressively flatten, eventually becoming
horizontal.
Plotting $S^{1.5}P_S$, we show the departure of $P_S$
from the MF expectation $P^{\tt MF}_S \propto S^{-1.5}$ as the critical point is approached.
When investigating the distribution of stress fluctuations 
$\eta_i = \sum_{j\neq i} G_{ij} \frac{n_{j}\sigma_{j}}{\tau}$ on each site, we find
consistently a change from a peaked distribution with fat tails towards Gaussian-like distributions
as we increase the shear-rate.
We infer from this, that the strong correlations at vanishing shear-rates
(reason for the non-trivial criticality) become negligible for stronger driving,
so that the exponents end up being well described by mean-field assumptions. 

(II) The second consequence of a finite driving rate is that the critical scaling regime shows
not only finite size, but also finite shear-rate effects~\cite{LemaitrePRL2009, KarmakarPRE2010}. 
When imposing a finite deformation rate, each stress-drop is characterized not only by
its magnitude or size $S$, but also by its duration $T$.
For each stress-drop we define a given duration $T$, as the time elapsed between the beginning
and the end of the drop.
In Fig.\ref{fig:FlowcurvePofXSofTPofT}(c) we present the distributions of durations $P_T$ for a
fixed system size and different shear-rates.
In the probed shear-rate regime we find the dependence on $L$ to be negligible,
thus $P_T(T,L,\gdot)\equiv P_T(T,\gdot)$.
The main panel shows rescaled curves assuming the functional dependence
$P_T \sim T^{-\tau'}g(T\gdot^{\alpha})$, with $g$ an exponentially decaying function.
We obtain for the $3d$ case, $\tau_{\tt 3d}'=1.44$ and $\alpha_{\tt 3d}=0.3$.
Naturally, we expect the scaling of $P_T$
to be dominated by a growing
length scale $\xi$ in the critical limit, where the relations $T\sim\xi^z$ and $S\sim\xi^{\df}$ hold.
Therefore, we expect a scaling relation $S \sim T^\delta $ with $\delta=\df/z$, that we observe
over a range of shear-rates, yielding the exponent $\delta_{\tt 3d} \sim 1.58$ (see Fig.\ref{fig:FlowcurvePofXSofTPofT}(d)),
in contrast with the mean-field $\delta_{\tt MF}=2$.
More generally, we observe empirically a power-law scaling of $S$ with $T$, $\gdot$ and $L$.
Actually, extending the dependencies of the cut-off values in size, $L^{\df}$, and duration, $\gdot^{-\alpha}$,
the mean $S$ at each $T$ should follow $\bar{S}(T,L,\gdot) = C(L,\gdot)T^{\delta}$ with $C(L,\gdot) \sim L^{\df}\gdot^{\alpha \delta}$.
This relation is fairly verified for the dependence on $\gdot$, illustrated in Fig.\ref{fig:FlowcurvePofXSofTPofT}(d).
A rescaling of the size dependence leads to an exponent larger by $15\%$ than $\df$ estimated from $P_S$.

\iffigures
\begin{figure}[t!]
\begin{center}
\includegraphics[width=\columnwidth, clip]{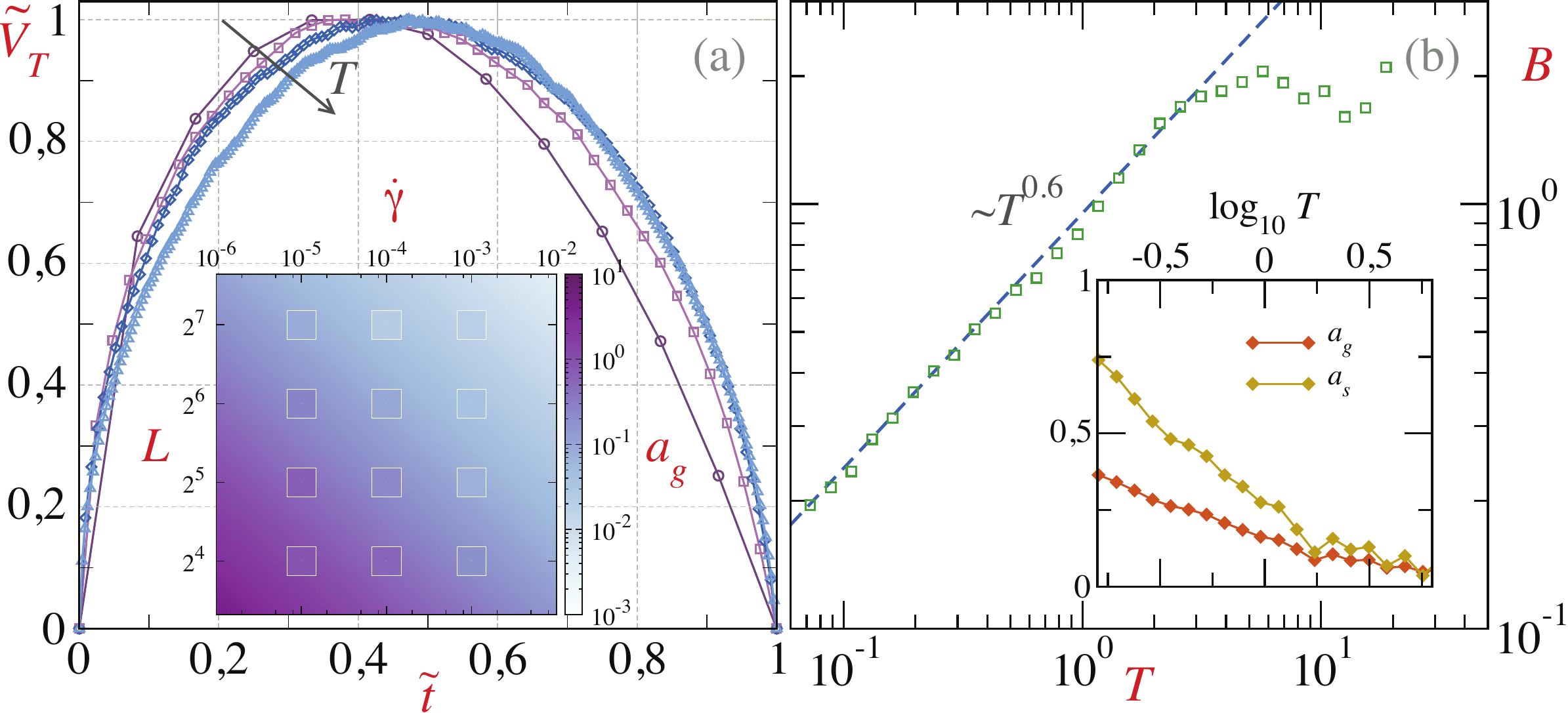}
\end{center}
\caption{\textit{Stress-drop shape properties for a 3$d$EP model}:
(a) Rescaled stress-drop shape $\tilde{V}_T(t)=V_T(t)/\max_{t}(V_T(t))$ averaged over
stress-drops of duration $T\pm\epsilon$, as a function of rescaled time $\tilde{t}=t/T$.
From left to right, we show curves at increasing $T$.
Inset: Bordered squares represent fitted values of the asymmetry parameter $a_g$ for
different choices of $(\gdot,L)$.
Color code depicts the fit $a_g=10^{-0.42}T^{-0.43}\gdot^{-0.37}L^{-1.25}$ for $T=0.5$.
(b) Amplitude $B$ of the stress-drops vs. $T$, for $L=32$ and $\gdot=10^{-4}$, as obtained from the fits.
The inset shows corresponding $a_s(T)$ and $a_g(T)$ (see text).
}
\label{fig:ShapesAndAsym}
\end{figure}
\fi

\noindent
{\it Stress-drop shapes--}
We address now the analysis of the functional form of the stress-drops, i.e.,
the time evolution of the stress-drop velocity~\cite{PapanikolauNatPhys2011,DahmenNatPhys2011,LaursonNatCom2013,AntonagliaPRL2014}.
In Fig.\ref{fig:ShapesAndAsym}(a) we show rescaled stress-drop velocities $V_T$
(stress-drop shapes) for a $3d$ system, averaged over drops of the same duration $T$
within the power-law scaling regime of Fig.\ref{fig:FlowcurvePofXSofTPofT}(d).
We observe that drops of short duration show a noticable asymmetric shape, with faster velocities at
earlier times.
As duration increases, the shape becomes gradually more symmetric.
To analyze this asymmetry of stress-drop shapes for different durations, system sizes and applied shear-rates,
we fit them with a formula proposed in Ref.\cite{LaursonNatCom2013}
$V_T(\tilde{t}) \propto B(\tilde{t}(1-\tilde{t}))^{c}(1-a_s(\tilde{t}-0.5))$ (see also~\cite{PapanikolauNatPhys2011,LeDoussalEPL2012}),
with $B$ the amplitude of the shape and $a_s$ a parameter quantifying the deviation from a symmetric inverted parabola.
We confirm the expected relation $c=\delta-1$ (recall $S \propto T^\delta$ and compare Fig.\ref{fig:ShapesAndAsym}(b)and Fig.\ref{fig:FlowcurvePofXSofTPofT}(d)). 
In our range of parameters $c$ is almost independent of $L$ and $\gdot$.
More relevant for our analysis is the behavior of the fitting parameter $a_s$
(see Fig.\ref{fig:ShapesAndAsym}(b) inset), that shows clearly the crossover from nearly symmetric
to asymmetric shapes as we focus on shorter durations $T$.
To avoid a fit with various parameters, 
we use an alternative, purely geometrical measurement of the asymmetry that 
is relevant even beyond scaling regime,
$a_g=\int_{0}^{1}\frac{|V_T(\tilde{t})-V_T(1-\tilde{t})|}{V_T(\tilde{t})+V_T(1-\tilde{t})}d\tilde{t}$.
When computing $a_g(T)$ for different shear-rates at fixed $T$ and $L$, $a_g$ increases as $\gdot$ decreases; 
whereas for fixed $T$ and $\gdot$, $a_g$ decreases as $L$ increases (see inset of Fig.\ref{fig:ShapesAndAsym}(a)).
In the quasistatic limit, where just one independent avalanche occurs at a time 
we expect asymmetric stress-drop shapes characterizing individual avalanches.
When we increase the driving rate at fixed system size or, equivalently
increase the system size at a fixed rate, we expect stress-drops 
to result from many independent avalanches, since
the density of plastic regions is determined
and increased by the driving strength~\cite{LemaitrePRL2009}.
Here, the resulting stress-drop shape draws closer to the mean-field symmetric shape.

\noindent
{\it Conclusions--}
We studied with a mesoscopic model the avalanche statistics close to the yielding transition,
verifying the relevance of our approach by comparing with particle-based quasistatic simulations.
In Table \ref{table:criticalexponents} we summarize the critical exponents obtained for $2d$ and $3d$.
Our results clearly reinforce the idea of a non-trivial universality 
class for the yielding transition, in agreement with earlier findings~\cite{SalernoRoPRE2013,TalamaliPRE2011,LinPNAS2014}.
Our estimated exponents, confirm within error bars the scaling relations proposed by Lin et \textit{al.}~\cite{LinPNAS2014}.
We also note that our values of $\tau$ and $\tau'$ are indistinguishable from the exponents expected for the
$1d$ long-range ($1/r^2$) depinning universality class~\cite{BonamyPRL2008, LaursonPRE2010}.
Although the loading path dependence of the critical exponents remains an open issue,
this is an interesting accordance and points towards the role played by the avalanche slip-line geometry.

In the regime of larger shear-rates we find that several exponents of the stress-drop statistics draw closer to mean-field predictions.
The rise of an increasing number of independent regions with yielding activity (parallel
occurring avalanches) justifies the crossover to trivially random statistics.
In particular our data reveals a yielding exponent approaching 
the prediction of the H\'ebraud-Lequeux model~\cite{HebraudPRL1998, AgoritsasEPJE15, PuosiArxiv15}.
Further the finite shear-rate protocol allows for the introduction of an additional exponent $\alpha$ 
that should enter the scaling relations.
If we assume a usual scaling scenario, we expect a diverging length scale
depending on the distance to the yielding point $\xi\sim (\sigma-\sigma_c)^{-\nu}$,
such that $\xi\sim\dot{\gamma}^{-\nu/\beta}$, since $\dot{\gamma}\sim(\sigma-\sigma_c)^{\beta}$.
Then $T\sim\xi^z$ yields directly the scaling relation $\alpha=z\nu/\beta$.
We have not measured $\nu$, but assuming 
$\nu=1/(d-\df)$~\cite{LinPNAS2014} to be valid we get $\alpha_{2d}=0.34$ and $\alpha_{3d}=0.31$, close to the measured values.

\begin{table}[t]
\begin{center}
 \begin{tabular}{l | c | c c | c | c }

  {} & {Expression} &  \multicolumn{2}{c|}{This work (2$d$ $|$ 3$d$)} & {\textit{lr}-depinning 1$d$} & MF \\
	\hline
	$\beta$ & $ \dot{\gamma}\sim(\Delta\sigma)^\beta$ & $1.54(2)$ &  $1.55(2)$  & $0.625(5)$~~\cite{DuemmerJStatMech2007}           & 2~~\cite{HebraudPRL1998} \\
	$\tau$  & $P_S\sim S^{-\tau}$                     & $1.28(5)$ &  $1.25(5)$  & $1.25(5)$~~\cite{LaursonPRE2010,BonamyPRL2008}  & 1.5~~\cite{DahmenNatPhys2011} \\
	$\df$   & $S_c \sim L^{\df}$                      & $0.90(7)$ &  $1.3(1)$   & $\sim1.38 $~~\cite{DuemmerJStatMech2007}            & --- \\
	$\tau'$ & $P_T\sim T^{-\tau'}$                    & $1.41(4)$ &  $1.44(4)$  & $\sim1.43$~~\cite{BonamyPRL2008}                 & 2~~\cite{DahmenNatPhys2011} \\
	$\alpha$& $T_c \sim \dot{\gamma}^{\alpha}$        & $0.38(4)$ &  $0.30(4)$  & ---                                              & --- \\
	$z$     & $T \sim \ell^z $                        & $\sim0.57$&  $\sim 0.82$& $0.77(1)$~~\cite{DuemmerJStatMech2007}             & --- \\
	$\delta$& $S \sim T^\delta$                       & $1.58(7)$ &  $1.58(5)$  & $\sim1.7$~~\cite{BonamyPRL2008}                 & 2~~\cite{DahmenNatPhys2011} \\
	$\theta$& $P_x \sim x^\theta$                     & $0.52(3)$ &  $0.37(5)$  & 0                                                & 1~~\cite{LinEPL2014} \\  
	\hline
 \end{tabular}
\caption{\label{table:criticalexponents} Measured exponents for the avalanche statistics. 
}
\end{center}
\end{table}

Within the scaling regime for $T$ we observe both asymmetric and symmetric stress-drop shapes
depending on system size, shear-rate and duration. 
This is why we propose to  distinguish between individual avalanches (resulting from correlated plastic events) 
and stress-drop shapes (resulting from many independently occurring avalanches).

The combined study of avalanche size and duration distributions and avalanche shapes has played an essential
role in our understanding of the universal aspects of crackling noise and depinning dynamics.
With this work, we provide a first numerical prediction of similar quantities in the case of the yielding
transition, with a clear indication of a complex non mean-field behavior.
We hope this work will stimulate and provide a benchmark for future experimental studies on systems
undergoing a continuous yielding transition, for which detailed data on noise statistics is presently very scarce. 

\begin{acknowledgments}
JLB, EEF and CL acknowledge financial support from ERC grant ADG20110209.
JLB is supported by IUF.
KM acknowledges financial support from grant ANR-14-CE32-0005 (project FAPRES).
EEF and JLB acknowledge the hospitality of the KITP, supported in part by the
National Science Foundation under Grant No. NSF PHY11-25915.
Most of the computations were performed using the Froggy platform of the 
\href{https://ciment.ujf-grenoble.fr}{CIMENT} infrastructure supported by the Rh\^one-Alpes region (GRANT CPER07-13
\href{http://www.ci-ra.org/}{CIRA}) and the Equip@Meso project (reference ANR-10-EQPX-29-01).
Further we would like to thank Alexandre Nicolas, Elisabeth Agoritsas, Eric Bertin, Jordi Ort\'in and St\'ephane Santucci
for fruitful discussions, and Mark Robbins and Matthieu Wyart for a useful correspondence.
\end{acknowledgments}

%

\end{document}